\documentclass[twocolumn,prl,aps]{revtex4}
\usepackage{graphicx}                                   
\tighten
\renewcommand{\vec}[1]{{\mathbf{#1}}}
\def\tm{T_{-1}}
\def\tone{T_1}
\def\tzero{T_0}
\def\resol{\frac{1}{E-V-T_0}}
\newcommand{\beq}{\begin{eqnarray}}
\newcommand{\eeq}{\end{eqnarray}}
\begin{document}
\draft
%\twocolumn
%\preprint{dvi file made on \today}
%\input epsf.sty
%\input psfig.sty

\title
{Absence of Asymptotic Freedom in Doped Mott Insulators:  Breakdown of Strong Coupling Expansions}
\author{  Philip Phillips and Dimitrios Galanakis}
\affiliation{Loomis Laboratory of Physics,
University of Illinois at Urbana-Champaign,
1110 W.Green St., Urbana, IL., 61801-3080}
\author{Tudor D. Stanescu}
\affiliation{Department of Physics and Astronomy, Rutgers University,
136 Frelinghuysen Rd., Piscataway, NJ., 08854-8019}
\begin{abstract}
We show that doped Mott insulators, such as the copper-oxide superconductors, are asymptotically slaved in that the quasiparticle weight, $Z$, near half-filling depends critically on the existence of the high energy scale set by the upper Hubbard band. In particular, near half filling, the following dichotomy arises: $Z\ne 0$ when 
the high energy scale is integrated out but $Z=0$ in the thermodynamic limit
when it is retained.  Slavery to the high energy scale arises from quantum
 interference between electronic excitations across the Mott gap.  Broad spectral features seen 
in photoemission in the normal state of the cuprates are argued to arise from high energy slavery.
\end{abstract}

\maketitle

Vast progress in theoretical solid state physics has been made by constructing  models which describe the low-energy properties of solids.  Essential to
the success of this program is the separability of the high and low energy degrees of freedom.  Should this separability hold, then the high energy scales can
 be integrated out yielding an effective Hamiltonian which describes the relevant low-energy or long-wavelength physics.  Notable successes 
include the prediction that dilute magnetic moments are quenched at low temperatures in non-magnetic metals (the Kondo problem)\cite{phillips} and the Landau Fermi liquid theory\cite{shankar} of the normal state of metals.

In the context of high-temperature superconductivity in the copper oxide materials, a similar approach has been adopted\cite{red1,red2,red3,red4,red5,red6,red7,red8}.   The high energy scale in the cuprates corresponds to the energy cost to doubly occupy the same `lattice site' in the copper-oxide plane.  At half-filling, an insulating state (the Mott insulator) obtains when the double occupancy energy cost ($U$) vastly exceeds the nearest-neighbour hopping energy, $t$.  In the cuprates,  $U\approx 10t$. This corresponds to the strong coupling regime.  In this regime, a requirement for any low-energy theory is that all hopping processes preserve the number of doubly occupied sites\cite{kohn}.  The resultant
Hamiltonian can then be projected onto the singly occupied subspace or lower Hubbard band and then studied
 accordingly\cite{red1,red2,red3,red4,red6,red7,red8}. 
Through second order in $t/U$, the spin-spin Hamiltonian (or $t-J$ model) that obtains has been used widely as an effective model for the cuprates\cite{red1,red2,red3,red4,red5}. In this Letter, we show that this procedure fails for the high-temperature superconductors because in the strong-coupling regime, the value of the quasiparticle weight, in the vicinity of half-filling, depends crucially on the presence of the upper Hubbard band. Namely models in which it is integrated out yield well-defined quasiparticles, whereas those in which it is not yield a vanishing weight in the thermodynamic limit.  At work here is a {\it non-perturbative} quantum interference effect between excitations that live at energy scales that span the Mott gap.

Our conclusion that the physics of projected models can be different from that of the Hubbard model
 suggests that the high and low energy degrees of freedom are coupled. 
Indeed, numerous experiments support this view. For example, optical conductivity measurements\cite{cooper,uchida1} as well as oxygen K-edge photoemission\cite{chenbatlogg} indicate that there is a massive re-shuffling of spectral weight from an energy scale as high as 2eV above the Fermi energy in both hole and electron-doped cuprates as a function of doping such that the
low-energy spectral weight (LESW) increases at the expense of the high-energy ($>2eV$)
spectral weight. Similar results are also seen in angle resolved photoemission\cite{ncco}. The non-trivial sum rule\cite{chenbatlogg,sawa} that has emerged from oxygen K-edge x-ray studies is that at least one single particle state (per doped hole) is
lost at high energies and transferred to low energies such that the LESW increases faster than $2x$, where $x$ is the hole-doping level.   The $t-J$ model does not preserve\cite{sawa} this sum rule as the LESW is exactly
$2x$ in this truncated scheme.  LESW in excess of $2x$ is purely dynamical and arises from virtual excitations to the upper Hubbard band. Such
spectral weight transfer indicates that some low-energy 
degrees of freedom in the normal state of the cuprates are derived from a high energy scale.  The strong electron correlations that give rise to a mixing between the low and high energy degrees of freedom in doped Mott insulators we termed {\it Mottness}\cite{stanescu}.  A further surprise\cite{rubhaussen,marel,bontemps} is that Mottness
persists even when superconductivity obtains.  For example,  R\"ubhaussen, et. al.\cite{rubhaussen} have shown
that changes in the optical conductivity occur at energies 3eV (roughly 
100$\Delta$, $\Delta$ the maximum superconducting gap) away from the 
Fermi energy at $T_c$, and Bolegr\"af, et. al.\cite{marel} have seen an 
acceleration in the depletion of the high energy spectral weight accompanied 
with a compensating increase in the low-energy spectral weight at and below the
superconducting transition.  Similarly, Bontemps, et. al.\cite{bontemps}
observed that
in underdoped (but not overdoped) cuprates, the Glover-Ferrel-Tinkham sum rule is violated and the optical
conductivity must be integrated to $\approx 100\Delta$, as opposed to $4\Delta$ in conventional superconductors, for the spectral
weight lost upon condensation into the superconducting state to be recovered.

Nonetheless, the experiments\cite{cooper,uchida1,chenbatlogg,ncco,rubhaussen,marel,bontemps} which demonstrate that all energy scales are mixed in the cuprates have had virtually no theoretical impact.  What seems to be missing is an explicit demonstration that a key physical quantity differs once the high energy scale is integrated out.  We propose that
a crucial quantity
that captures the difference between projected models and the Hubbard model is the quasiparticle weight. 

To proceed, we first show that all problems regarding the formulation of projected models to high order can be overcome.  Hence, should a failure arise, it does not reside in the formulation.  Consider the simplest model for a doped Mott insulator.  In the Hubbard model, 
\beq
H=T+V=-t\sum_{\langle ij\rangle} c_{i\sigma}^\dagger c_{j\sigma}+U\sum_i n_{i\uparrow}n_{i\downarrow}
\eeq
electrons acquire kinetic energy by hopping among neighbouring sites, $\langle ij\rangle $ and experience an on-site repulsive interaction, $U$.  It is expedient to break the kinetic energy into three terms,
\beq
T_0&=&-t\sum_{\langle ij\rangle} \left(\eta^\dagger_{i\sigma}\eta_{j\sigma}+\xi^\dagger_{i\sigma}\xi_{j\sigma}\right),\nonumber\\
T_1&=&-t\sum_{\langle ij\rangle}\eta^\dagger_{i\sigma}\xi_{j\sigma},\nonumber\\
T_{-1}&=&-t\sum_{\langle ij\rangle}\xi^\dagger_{i\sigma}\eta_{j\sigma},
\eeq
which are eigenoperators of the interaction and as a result, obey the
 commutator, $[V,T_m]=mUT_m$.  The operator $T_m$ increases the double occupancy by $m$. The operators
$\eta_{i\sigma}=c_{i\sigma}n_{i-\sigma}$ and $\xi_{i\sigma}=c_{i\sigma}(1-n_{i-\sigma})$ annihilate electrons on doubly and singly-occupied sites, respectively. Note, $c_{i\sigma}=\eta_{i\sigma}+\xi_{i\sigma}$.  Successful removal of double occupancy implies that $\eta_{i\sigma}$ and $\xi_{i\sigma}$ can be decoupled.  For simplicity, we will set $U=1$.  

To show how such excitations enter the projected Hamiltonian schemes, we review the two standard perturbative approaches used in this context.   In the first approach to removing double occupancy, we use a similarity transformation, $S$, such that the transformed Hamiltonian 
\beq\label{htilde}
\tilde H=e^SH e^{-S}.
\eeq
does not contain hops between sites with differing numbers of doubly occupied sites.
As this procedure is well-described\cite{kohn,oles,takahashi,mgy} in the literature,
we will be brief.  Our Hamiltonian initially is 
$H=V+T_0+T_{-1}+T_{1}$.  The last two terms in the Hamiltonian do not conserve the number of doubly occupied sites and hence must be eliminated.  In the standard implementation, the 
similarity transformation\cite{mgy} is chosen such that
$T_{-1}+T_1+[V,S^{(1)}]=0$.
The transformation that accomplishes this is $S^{(1)}=T_1-T_{-1}$.   At each order, the similarity transformation must be modified accordingly.  To obtain the effective Hamiltonian in the singly
occupied subspace, we perform the projection $P_0\tilde{H}P_0$ which removes all terms in which $T_{-1}$ appears first as a consequence of $T_{-1}|0\rangle=0$, where $|\rm 0\rangle$ is any state in the lower Hubbard band. Through fourth order, we write 
the effective Hamiltonian\cite{mgy}, $H_{\rm eff}=H_{\rm red}+H_{\rm irred}$, as a sum of irreducible,
\beq
H_{\rm irred}&=&T_0-T_{-1}(1+\tzero+\tzero^2)T_1+\tm^2\tone^2
\eeq
and reducible,
\beq
H_{\rm red}=\{\tzero-\tzero^2,\frac{\tm\tone}{2}\}+\{\tzero,\tm\tzero\tone\}+(\tm\tone)^2\nonumber
\eeq
terms where $\{a,b\}=ab+ba$. Third and higher order terms are non-zero away from half-filling in bi-partite lattices.  Since each term preserves the number of doubly occupied sites, the sum of the indices on each product of $T_n$'s vanishes.  In the irreducible terms, all the intermediate states contain at least one doubly occupied site. The energy denominators ( the 1/U factors) arise from this energy difference.   The reducible terms are products of irreducible ones and hence they contain hopping processes that do not originate from excitation to the doubly occupied subspace, $T_{-1}T_1T_0$ nor terminate once an electron is returned to the
 singly occupied subspace, for example, $(\tm\tone)^2$.  All such processes can be viewed as arising from a rotation\cite{stein} of the eigenstates in the low-energy sector. Such a rotation arises naturally in this context, since a unitary transformation preserves orthogonality.  While it might be anticipated that the true effective low-energy Hamiltonian should be independent of such a rotation and hence irreducible with respect to the target manifold, such is not the case here. 
 In fact, the super-extensive parts of $H_{\rm irred}$ cancel those of $H_{\rm red}$, giving rise to a linked expansion for the energy.  The first two terms in $H_{\rm irred}$ yield the $t-J$ model in addition to the three-site hopping which describes the motion of a hole in a spin background.  

The relationship between the change of basis and size consistency is further
illustrated using 
Brillouin-Wigner (BW) perturbation theory. Let $P$ be the projector for the lowest degeneracy subspace and $Q=1-P$, the orthogonal complement. Because $V$ and $T_0$ do not change the number of doubly occupied sites,
$[Q,V+T_0]=0$.   Consider the Schr\"odinger equation, $(E-V-T)|\psi\rangle=0$, where $|\psi\rangle$ is the exact many-body eigenstate in the Hubbard model and $P|\psi\rangle=|\psi_0\rangle$ yields the exact eigenstate in the zero-double occupancy sector.  Multiplying
 the Schr\"odinger equation on the left by $Q$ results in the 
formal expansion
$(E-V-T_0)Q|\psi\rangle=Q(T_{-1}+T_1)|\psi\rangle$
for the components of the eigenstates orthogonal to those in the zero double occupancy subspace.To compute the energy eigenvalue, we multiply the Schr\"odinger equation on the left by $P$ to obtain
$E|\psi_0\rangle=T_0|\psi_0\rangle+PT_{-1} Q|\psi\rangle$.
Successively iterating this equation twice by using the equation for $Q|\psi\rangle$ yields the self-consistent expansion 
\beq
E|\psi_0\rangle&=&\left(T_0+P\tm\resol\tone+P\tm\resol\right.\nonumber\\
&&\left.\times\tm\resol\tone\resol\tone\right)|\psi_0\rangle
\eeq
for the energy through fourth order.  To obtain a more useful form for the energy eigenvalue, we expand the energy denominators and multiply on the left by $\langle\psi_0|$.  What results is a second-order polynomial
in $E$.  The unique root that vanishes as $t\rightarrow\infty$ is given by
\beq\label{energy2}
E=\langle H_{\rm irred}\rangle-\langle T_{0}- T_{-1}T_{1}\rangle\langle T_{-1}T_{1}\rangle+\langle T_0\rangle^2\langle T_{-1}T_{1}\rangle
\eeq
through $O(t^4)$.  The expectation value in Eq. (\ref{energy2}) is performed with the exact eigenstates in the subspace with zero double occupancy.  Typically in degenerate perturbation theory, a basis which lifts the degeneracy to first order is sufficient to evaluate all the higher order terms.  In this case, this would correspond to using a basis that diagonalizes $T_0$.  In such a basis, $\tm\tone$ is not diagonal and hence the unlinked part of $\langle\tm\tm\tone\tone\rangle=\langle(\tm\tone)^2\rangle\propto N^2$ is not canceled. This problem is endemic to the Hubbard model, because in traditional perturbation theories there is no analogue of $T_0$ which induces transition only in the target space.  The correct scaling with $N$ is accomplished by expanding $|\psi_0\rangle$ in powers of $t$ and collecting all unlinked terms order by order in $t$.  If this is done, the unlinked part of the terms containing a single factor $T_0$ and $(\tm\tone)^2$ in $H_{\rm irred}$ cancel the second term in Eq. (\ref{energy2} and the unlinked part of the $\tzero^2$ terms in $H_{\rm irred}$ cancel the last term in Eq. (\ref{energy2}) through fourth order.  The unlinked parts of these terms of course have $O(t^5)$ contributions and higher.  All of these terms can be shown to cancel by the order by order expansion of the eigenstates in the lowest energy sector.  Consequently, perturbation theory and the canonical transformation will  be equivalent up to an arbitrary rotation in the target space.  While such rotations affect $H_{\rm red}$ not $H_{\rm irred}$, $H_{\rm red}$ is crucial to the correct size dependence of the true effective Hamiltonian.

\begin{figure}
\centering
\includegraphics[height=9.5cm]{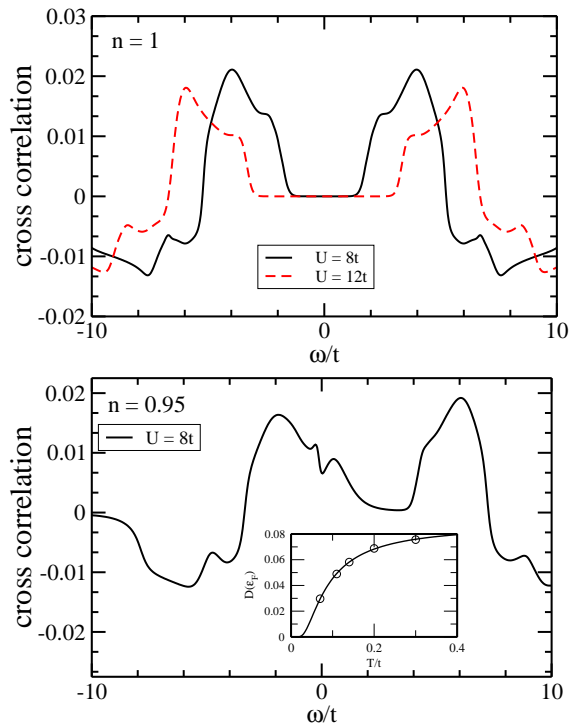}
\caption{Cross correlation or quantum interference $A_{\eta\xi}$ between the upper and lower Hubbard bands at half-filling, $n=1$ and at $n=.95$ at $T=0.1t$.
The dip at the chemical potential in the lower panel represents the pseudogap. The inset shows that this dip leads to a vanishing density of states at zero temperature and hence an orthogonality catastrophe.}
\label{crossterm}
\end{figure}

While the subtleties in constructing projected effective Hamiltonians can
be overcome up to an arbitrary rotation in the target space, all such expansions rely on the partitioning of the electron into $\xi_{i\sigma}$ and $\eta_{i\sigma}$ excitations.  Consequently, the full electron spectral function,
$A(\vec k,\omega)=-{\rm Im}FT(\theta(t-t')\langle \{c_{i\sigma}(t),c^\dagger_{j\sigma}(t')\}\rangle)/\pi= A_{\eta\eta}+A_{\xi\xi}+2A_{\eta\xi}$,
contains two diagonal terms corresponding to the upper and lower Hubbard bands, $A_{\eta\eta}$ and $A_{\xi\xi}$, respectively and a cross term $A_{\eta\xi}$ which represents the degree to which the high and low energy degrees of freedom are coupled.  Here, FT represents the frequency and momentum Fourier transform. Shown in Fig. (\ref{crossterm}) is an explicit calculation of the
$A_{\eta\xi}$ term (integrated over $\vec k$) using the dynamical two-site method detailed previously\cite{stanescu}.  As in other cellular methods\cite{cellular}, the self-energy for the lattice is constructed from the resolvents for the electronic states on a finite cluster using a self-consistent closure. In this case, a two-site cluster is used.   As is evident from Fig. (\ref{crossterm}), $A_{\eta\xi}$ is distinctly {\bf non-zero} and mirrors the overall single particle density of states with peaks at the upper and lower Hubbard bands.  Three features are most relevant. First, at $U=8t$, and at half-filling, the overall density of states (see Fig. 4 of Phys. Rev. B,  {\bf 64} 235117/1-8 (2001)) has a maximum value of $0.16$, whereas the total weight arising from the cross term, $2A_{\eta\xi}$, is $.04$ or 25\%.  Hence, this contribution cannot be ignored.  Second, the cross term has both negative and positive contributions. This structure arises necessarily because the integral of 
$A_{\eta\xi}$ over all frequency yields the equal time correlator 
$\langle\{\xi_{\i\sigma},\eta^{\dagger}_{i\sigma}\}\rangle=0$, whose vanishing maintains the Pauli principle. This implies that $A_{\eta\xi}$ is either zero, which it is not, or it must have both positive and negative parts, representing constructive and destructive interference, respectively, between different regions in energy space.  Third, when $U$ is increased by $50\%$ to $U=12t$ as in the cuprates, the cross term does not decrease appreciably.   Nonetheless, at half-filling, the contribution of the crossterm below and above the chemical potential sum to zero independently, indicating that the upper Hubbard band
can be integrated out safely without sacrificing the Pauli principle.

Such is not the case, however, at finite doping.  The lower panel in 
Fig. (\ref{crossterm}) indicates that a pseudogap develops at the chemical potential, indicating an orthogonality catastrophe and hence a vanishing of the quasiparticle weight.  Because a pseudogap subtracts spectral weight at low energies and transfers it to the upper Hubbard band, the sum rule which ensures the Pauli prinicple is satisfied  only when $A_{\eta\xi}$ is integrated over all energy scales not simply up to the chemical potential (or some intermediate energy cutorr) as would be case in projected models.  Hence, although the cross term {\it can} be obtained perturbatively from projected schemes by canonically transforming (as described
 previously) the electron operators, perturbation theory fails as the {\it integrated} cross term represents inherently {\it non-perturbative} physics, namely the Pauli principle.  Symptomatic of this failure is the difference in the value of the single-hole quasiparticle weight, $Z=|\langle \psi_G|c_{k\sigma}^\dagger|\psi_{\vec k,\sigma}\rangle|^2$ between the Hubbard and projected models.  Here  $\psi_G$ and $\psi_{\vec k,\sigma}$ are the exact ground states for the half-filled and one-hole systems, respectively.  In projected schemes, such as the $t-J$ model, $Z\propto J/t$ 
as has been demonstrated both analytically and numerically\cite{dagotto}. Consequently, a single hole is delocalized at $T=0$ in the t-J model.  However, in the Hubbard model, adding a single hole leads to a ``non-renormalizable'' phase shift of each state in the first Brillouin zone and hence an orthogonality catastrophe\cite{red1,sorella} in the thermodynamic limit,
 $Z\propto L^{-\delta}$\cite{red1,sorella}. Consequently,
 $Z$ for one-hole in the Hubbard model at half-filling does not appear to have a well-defined expansion in $t/U$.  It is this breakdown that we term asymptotic slavery.  This failure, applies strictly in the thermodynamic limit and should persist as long as a pseudogap is present which necessarily leads to a vanishing of the quasiparticle weight.  Broad spectral features
\cite{ncco} (that is, $Z=0$), spectral weight transfer\cite{chenbatlogg}, hole localization in the underdoped regime,  as well as the color change seen in optical experiments\cite{marel} upon a transition to the superconducting state are all signatures of the quantum interference that is the root of asymptotic slavery.

Of course, asymptotic slavery in doped Mott insulators stands in stark contrast to the perturbative physics present at short distances, that is asymptotic freedom, in quark matter\cite{gross,politzer}.  Short of an exact construction of the quasiparticles, any realistic model of the cuprates must be solved on the energy scale $U$ because double occupancy does not necessarily mean high energy. That is, a Wilsonian renormalization group analysis fails as long as asymptotic slavery is present, namely as long as the pseudogap presists. The high energy scale does not simply renormalize the low-energy degrees of freedom.  In doped Mott systems, the Pauli principle appears as a sum rule over high and low energies.  Methods\cite{stanescu} which emphasize local non-perturbative physics or perhaps non-commutative field theories\cite{noncom} (which display UV-IR mixing) have the ingredients to capture how Mottness conspires to
yield asymptotic slavery. 

\acknowledgements This work was supported by the NSF, Grant No. DMR-0305864 and the ACS Petroleum Research Fund. We also thank A. Chernyshev, E. Fradkin, S. Girvin,
 G. Murthy, A. Rozhkov, R. Shankar, Q. Si, , A.-M. Tremblay, and  P. Wolynes for insightful comments.

\end{document}